\documentclass[aps,prl,reprint,superscriptaddress]{revtex4-2}
\usepackage{graphicx}
\usepackage{textcomp}
\usepackage{textgreek}
\usepackage{amsmath}

\begin{document}

% Use the \preprint command to place your local institutional report
% number in the upper righthand corner of the title page in preprint mode.
% Multiple \preprint commands are allowed.
% Use the 'preprintnumbers' class option to override journal defaults
% to display numbers if necessary
%\preprint{}

%Title of paper
\title{Optical control of the spin-Hall effect in a two-dimensional hole gas}

% repeat the \author .. \affiliation  etc. as needed
% \email, \thanks, \homepage, \altaffiliation all apply to the current
% author. Explanatory text should go in the []'s, actual e-mail
% address or url should go in the {}'s for \email and \homepage.
% Please use the appropriate macro foreach each type of information

% \affiliation command applies to all authors since the last
% \affiliation command. The \affiliation command should follow the
% other information
% \affiliation can be followed by \email, \homepage, \thanks as well.
\author{Simone Rossi}
%\email[]{Your e-mail address}
%\homepage[]{Your web page}
%\thanks{}
%\altaffiliation{}
\affiliation{Dipartimento di Scienza dei Materiali, Universit\`a degli Studi di Milano-Bicocca and BiQuTe, Via R. Cozzi 55, 20125 Milano, Italy}

\author{Valentina Caprotti}
\affiliation{Dipartimento di Scienza dei Materiali, Universit\`a degli Studi di Milano-Bicocca and BiQuTe, Via R. Cozzi 55, 20125 Milano, Italy}

\author{Andrea Filippi}
\affiliation{Dipartimento di Scienza dei Materiali, Universit\`a degli Studi di Milano-Bicocca and BiQuTe, Via R. Cozzi 55, 20125 Milano, Italy}

\author{Emiliano Bonera}
\affiliation{Dipartimento di Scienza dei Materiali, Universit\`a degli Studi di Milano-Bicocca and BiQuTe, Via R. Cozzi 55, 20125 Milano, Italy}

\author{Jacopo Pedrini}
\affiliation{Dipartimento di Scienza dei Materiali, Universit\`a degli Studi di Milano-Bicocca and BiQuTe, Via R. Cozzi 55, 20125 Milano, Italy}

\author{Roberto Raimondi}
\affiliation{Dipartimento di Matematica e Fisica, Universit\`a Roma Tre, Via della Vasca Navale 84, 00146 Roma, Italy}

\author{Maksym Myronov}
\affiliation{Department of Physics, The University of Warwick, Gibbet Hill Road, CV4 7AL Coventry, UK}

\author{Fabio Pezzoli}
\email[]{fabio.pezzoli@unimib.it}
%\homepage[]{Your web page}
%\thanks{}
%\altaffiliation{}
\affiliation{Dipartimento di Scienza dei Materiali, Universit\`a degli Studi di Milano-Bicocca and BiQuTe, Via R. Cozzi 55, 20125 Milano, Italy}

%Collaboration name if desired (requires use of superscriptaddress
%option in \documentclass). \noaffiliation is required (may also be
%used with the \author command).
%\collaboration can be followed by \email, \homepage, \thanks as well.
%\collaboration{}
%\noaffiliation

%\date{\today}

\begin{abstract}
Relativistic effects influence the motion of charged particles in solids by intertwining spin and momentum. The resulting phenomena exhibit rich and intriguing properties that can unveil radically new quantum devices. In this context, the two-dimensional hole gas formed in group IV heterostructures is a particularly promising platform, owning to a notable spin-orbit coupling. However, the exploitation of spin-momentum locking and precise manipulation of spin currents has remained elusive thus far. Here we use the modulation-doping technique to break inversion symmetry at novel $\textrm{Ge}_{\textrm{1-x}}\textrm{Sn}_{\textrm{x}}$/Ge interfaces and explore spin-orbit phenomena in the emergent Rashba-coupled hole gases. Magneto-optical investigations demonstrate the unusual establishment of a staggered band alignment with carrier lifetime in the ns range. Optical spin orientation is then leveraged to directly inject spin-polarized currents in the Rashba-split 2D gas. Spin-to-charge conversion is shown to genuinely occur at the staggered gap through the inverse spin-Hall effect. This provides unprecedented access to low-order contributions of the spin-orbit Hamiltonian. Moreover, it leads to the startling demonstration that the spin Hall angle can be optically controlled by modifying the Rashba coupling through the photoexcitation density. $\textrm{Ge}_{\textrm{1-x}}\textrm{Sn}_{\textrm{x}}$ quantum wells thus offer innovative solutions and functionalities stemming from their unique spin-dependent properties and intriguing quantum phenomena at the crossroad between transport and photonic realms.
\end{abstract}

% insert suggested keywords - APS authors don't need to do this
%\keywords{}

%\maketitle must follow title, authors, abstract, and keywords
\maketitle

% body of paper here - Use proper section commands
% References should be done using the \cite, \ref, and \label commands
\section{Introduction}\label{sec1}
Spin-orbit interaction is a relativistic effect that couples spin and momentum of charged particles. In solids this gives rise to the emergence of rich and intriguing spin-dependent phenomena, whose exploitation can advance the thriving field of quantum technologies.\cite{soumyanarayanan16, chatterjee21, trier22} Group IV materials may unleash tremendous opportunities in this context being a leading-edge platform perfected through decades of nanotechnology research. 

Specifically, Ge-based heterostructures can confine electrons in the form of a two-dimensional (2D) Fermi system, offering an ideal playground to tailor functional properties through spin-orbit coupling (SOC). This has been exemplified by the demonstration of long spin relaxation time and large anisotropy of the electron Land\'e g-factor in Ge/SiGe quantum wells (QW).\cite{giorgioni16} The formation of a 2D hole gas (2DHG) in Ge QWs has also led to the demonstration of g-factor manipulation \cite{mizokuchi18, myronov2023b} and exceptionally high mobilities, i.e., exceeding $10^6$ $\mathrm{cm^2/Vs}$. \cite{Myronov23} Recently it  has been utilized as a primer for the fabrication of hole spin qubits relevant for quantum information processing.\cite{hendrickx21,scappucci21} Indeed, the strong SOC that pertains to the valence band allows fast quantum logic operations. At the same time, hyperfine interactions are minimized by the \emph{p}-type orbital character of the hole wavefunction, thus offering a long coherence time.\cite{braakman21,hendrickx20}

While the introduction of an electrostatic potential through modulation doping (MOD) or a gate lead traps the 2DHG at the Ge/SiGe interface, it spontaneously breaks the structural symmetry of the system.\cite{manchon15} This notably generates an unusual spin texture, which stems from Rashba spin splitting.\cite{bihlmayer15, Moriya14, Morrison14, Zhao20} The latter might offer a compelling platform to develop persistent spin helix states that can be used to implement spin interconnects or quantum buses using the state-of-the-art manufacture of integrated circuits.\cite{koralek09, dery11} Above all, such Rashba-split hole gas provides a favorable testbed to explore synthetic spin-orbit fields that sustain spin-polarized states even at zero magnetic field. These are of primary interest in the quest to control the orbital motion of carriers and to manipulate the mutual conversion between spin and charge currents via the spin galvanic (Edelstein) and spin Hall effects.\cite{Dyakonov61, dyakonov71, ivchenko1978, Hirsch99} A context in which 2DHGs based on group IV materials remain vastly untapped.
 
In this scenario, the novel class of group IV alloys based on the heavy element Sn holds excellent spin properties.\cite{decesari19, vitiello20, Ferrari23} The investigation of SOC manipulation in $\mathrm{Ge_{1-x}Sn_x}$ heterostructures offers indeed the unique possibility to judiciously modify the system Hamiltonian by strain and bandgap engineering, thus providing radically new possibilities to implement spin-dependent functionalities beyond Ge. In addition, we anticipate that the strong light-matter interaction pertaining to Sn-based alloys \cite{decesari19, moutanabbir21} and the flexible design introduced by SOC engineering can make $\mathrm{Ge_{1-x}Sn_x}$ 2DHG a prominent candidate also at the leading edge between Si photonics and the burgeoning field of spin-orbitronics.\cite{soumyanarayanan16}  

In this work we leverage optical spectroscopy to explore these advanced capabilities. Specifically, we investigate the spin physics of individual \emph{p}-type MOD QWs realized by embedding $\mathrm{Ge_{0.91}Sn_{0.09}}$ layers within barriers made of elemental Ge. The formation of a 2DHG at cryogenic temperatures is proved by magneto-transport measurements, while excitation-dependent photoluminescence (PL) provides us with the fingerprint of an unexpected staggered band lineup for \emph{L}-valley electrons. A finding that advances our understanding of the electronic structure of these novel heterojunctions. Optical spin orientation is then exploited, in conjunction with external magnetic fields, to determine the spin lifetime through the Hanle effect.\cite{vitiello20, Dyakonov84} Notably, we observe spin-to-charge interconversion as a manifestation of the inverse spin-Hall (ISHE) effect in dedicated Hall-bar devices.\cite{Wunderlich05, ando11, okamoto14, bottegoni14} A result that allow us to provide a first estimation of the spin-Hall angle pertaining to this 2D system. In particular, these investigations demonstrate that heterojunctions featuring a distinct staggered gap can offer convenient control through direct laser excitation of the spin-to-charge conversion efficiency. Optical pumping has been finally utilized to unveil Rashba terms in the spin-orbit Hamiltonian that are linear in momentum. Such low-order component remained concealed to recent experimental observations of 2D-SOC systems.\cite{Moriya14, Morrison14, Zhao20, tai21} 

These findings open new research directions for the generation of pure spin currents and introduce group IV heterostructures as a practical platform for the future implementation of spin-orbitronic and spin-optronic functionalities.

\section{Results and Discussion}\label{sec2}

\begin{figure*}
  \includegraphics[width=\linewidth]{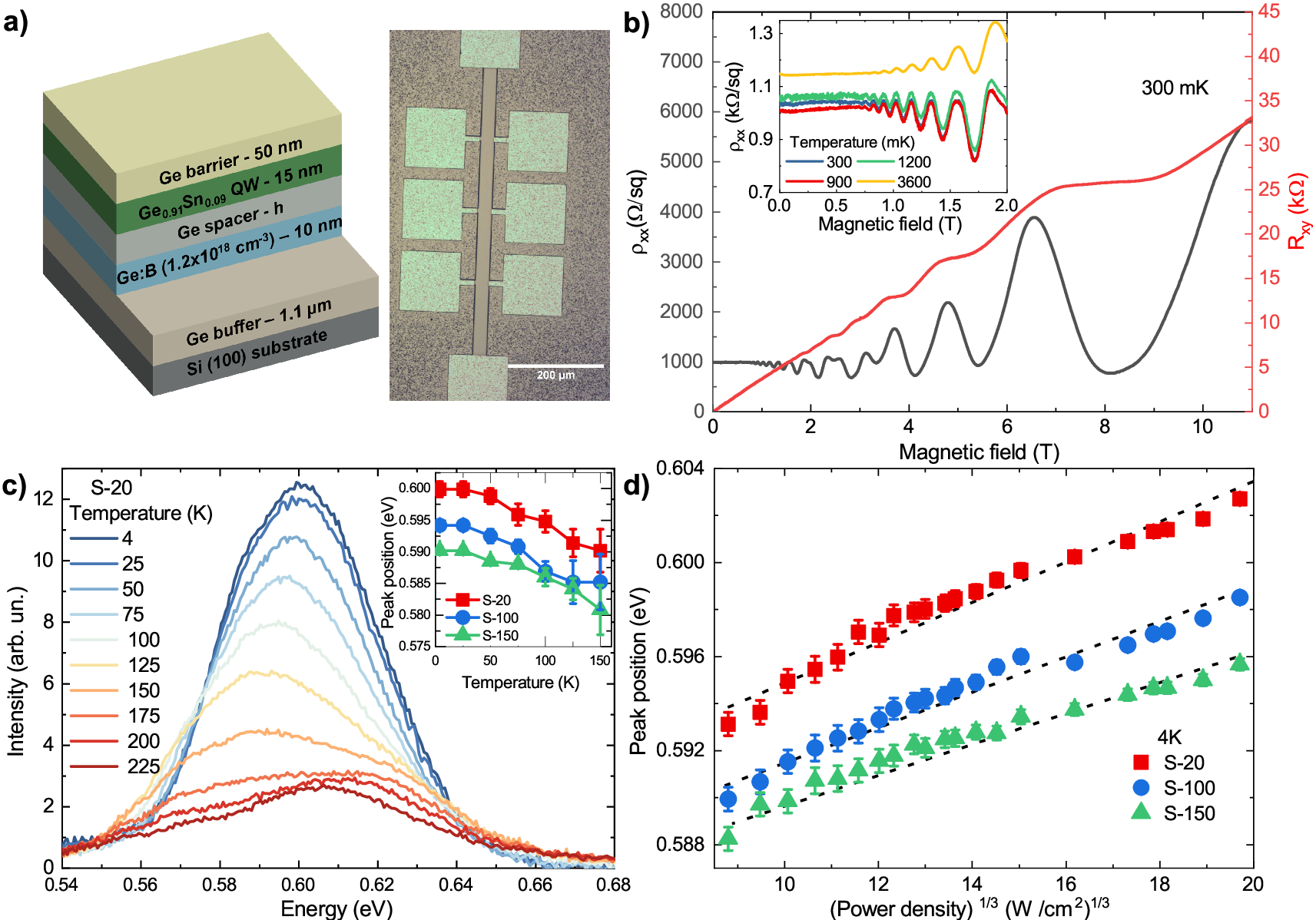}%
  \caption{(a) Schematic representation of the heterostructure. The Ge spacer thickness (h) is changed between the three samples: h = 20 nm for S-20, h = 100 nm for S-100 and h = 150 nm for S-150. On the right an optical image of a Hall bar structure patterned for transport and optical measurements. (b) Magnetotransport results at 300 mK showing Shubnikov-De Haas oscillations, i.e., the fingerprint of two-dimensional hole gas formation, as well as plateaus in the Hall resistance. The oscillations are found in the whole temperature range of 0.3 – 3.6 K. (c) Photoluminescence of S-20 in the temperature range 4 - 225 K under an illumination power density of 3.40 $\mathrm{kW/cm^2}$. (d) Peak position versus the third root of excitation power density showing a linear behavior.\label{fig1}}
\end{figure*}

\subsection{Electronic and optical properties of 2DHGs in $\mathrm{Ge_{1-x}Sn_x}$ QWs}

\textbf{Figure \ref{fig1}a} shows the cross-sectional layout of the epitaxial layers forming the \emph{p}-type MOD device along with an optical image of a representative geometry of the Hall bar used in this work. The thickness of the Ge spacer separating the $\mathrm{Ge_{0.91}Sn_{0.09}}$ QW from the Ge:B layer is of about 20 (S-20), 100 (S-100) or 150 (S-150) nm. See the Methods section for details about the sample growth and device fabrication. 

At first, magnetotransport investigations have been conducted through resistivity and Hall effect measurements at temperatures varying in the $0.3 – 3.6$ K range. The main results are shown in Figure \ref{fig1}b for sample S-20. While the transverse resistance ($R_{xy}$) shows well-defined plateaus at high magnetic fields, distinct Shubnikov-de Haas oscillations are clearly visible in the longitudinal resistivity ($\rho_{xx}$) over the whole temperature range, thus confirming the formation of a high-mobility 2DHG in the $\mathrm{Ge_{0.91}Sn_{0.09}}$ QW. The visibility of the oscillation pattern let us expect that the gas can withstand high temperatures operation. This is corroborated by results on similar structures where 2DHGs were singled out up to 5 K.\cite{gul20} At 300 mK a 2DHG carrier density of $1.7\times 10^{11} \mathrm{cm^{-2}}$ and mobility of $\approx 37300$ $\mathrm{cm^{2}V^{-1}s^{-1}}$ was derived, suggesting negligible conduction through parasitic channels. Notably, the biaxial compressive strain in the QW plane shifts the light-hole states at higher energies. This reduces the orbital mixing in the valence band and ensures that only the lowest heavy-hole (HH) subbands are populated.\cite{gul20, tai21} Finally, in Figure \ref{fig1}b a double peak can be observed to arise at 1.56 T ($\nu = 9$), which can be ascribed to the Zeeman spin splitting of the HH states.\cite{gul20}   

After having provided a compelling proof of the formation of the 2DHG through transport data, we are well positioned to focus on light-matter phenomena arising in this novel system. The excellent quality of the samples remarkably ensures that they are sufficiently bright to allow the direct observation of the radiative emission from the individual QW. This is readily demonstrated by the temperature-dependent photoluminescence (PL) summarized in Figure \ref{fig1}c.

It should be noticed that the strain imparted to the $\mathrm{Ge_{0.91}Sn_{0.09}}$ QW by the Ge-on-Si buffer has repercussions not only on transport, as previsoulsy discussed, but also on the PL. Beside lifting the valence band degeneracy, compressive strain strengthens the indirect character of the bandgap. Consequently, the peak appearing at cryogenic temperatures in Figure \ref{fig1}c at about 0.6 eV for sample S-20 can be ascribed to indirect transitions that involve the lowest subbands of the HH at the zone center and electrons at the $L$ point of the Brillouin zone.\cite{lin18, song19}  Owning to the Varshni's law, this spectral feature, termed $\mathrm{HH1-cL1}$, redshifts as the temperature increases. The same behavior holds for the samples S-100 and S-150 (see the inset of Figure \ref{fig1}c). 

$\mathrm{Ge_{1-x}Sn_x}$ heterojunctions have been developed only very recently as a means of band engineering, and various contrasting possibilities have been put forward thus far to describe offsets and discontinuities in the individual bands. Specifically, the conduction band lineup at $\mathrm{Ge_{1-x}Sn_x/Ge}$ heterointerfaces has remained undetermined with debated configurations that have been often derived from theoretical predictions, despite the lack of accurate calculation parameters. Presently, conflicting straddling (type I) \cite{gul20, Qian17, lin18} and staggered (type II) \cite{Stange16} alignments have been suggested. In the following, we demonstrate that optical spectroscopy can be instrumental to experimentally resolve such crucial ambiguities.

In type I heterostructures, like the prototypical $\mathrm{Ge/Si_{0.85}Ge_{0.15}}$ QWs,\cite{giorgioni16, kuo05, bonfanti08} electrons and holes are both confined in the same QW layer. Conversely, a type II band lineup as in Si/Ge QWs isolates carriers at opposite sides of the interface according to the sign of the charge,\cite{schaeffler97} thereby forming spatially indirect excitons alongside a dipole layer. The latter causes band-bending in the vicinity of the junction, whose steepness can be eventually modified through optical excitation. An increase of the pump power raises the carrier population, thus strengthening the electrostatic potential across the interface. This, in turns, enlarges the quantization energy so that a blueshift of the PL can be expected. The increase of the emission energy with the third root of the excitation density has been shown to provide a compelling proof of the establishment of such staggered lineup.\cite{Ledentsov95, Pavarelli13, Abramkin18, Timofeev18}

In the following, we utilize the spectral shift analysis to investigate the band alignment at the $\mathrm{Ge_{0.91}Sn_{0.09}/Ge}$ heterointerface. Figure \ref{fig1}d demonstrates that a pump-induced blueshift can be consistently observed for all the samples. Since many-body effects and heating from the laser excitation would have rather caused a redshift, owning to band-gap renormalization and temperature-induced bandgap narrowing, we can safely conclude that the unambiguous high-energy drift reported in Figure \ref{fig1}d demonstrates the emergence of state filling. This is further confirmed by the observation that the energy shift nicely scales with the third root of the excitation power. Such remarkable finding makes a strong case for a type II heterostructure, in which the 2DHG is confined in the $\mathrm{Ge_{0.91}Sn_{0.09}}$ QW and is surrounded by Ge barriers that become negatively charged under out-of-equilibrium conditions. 

It can be further noticed that at a fixed pump power, the samples exhibit an energy increase of the $\mathrm{HH1-cL1}$ peak as the thickness of the Ge spacer is reduced (see Figure \ref{fig1}d and the inset of Figure \ref{fig1}c). The reduction of the distance between the Ge:B doped region and the QW facilitates indeed trapping of the extrinsic charges by the well and increases the exerted electric field. This enlarges the density of holes that reside in the QW, progressively filling up the lowest energy states as the spacer is thinned down. This process forces photo-generated carriers to occupy higher energy levels, eventually yielding the observed blue-shift of the PL as a function of the width of the Ge spacer. This finding further supports the presence of a 2DHG in the QWs and suggest that both the MOD structure and the photoinduced state filling can contribute to the overall blueshift of the PL.

\subsection{Spin-resolved properties}

\begin{figure*}
  \includegraphics[width=\linewidth]{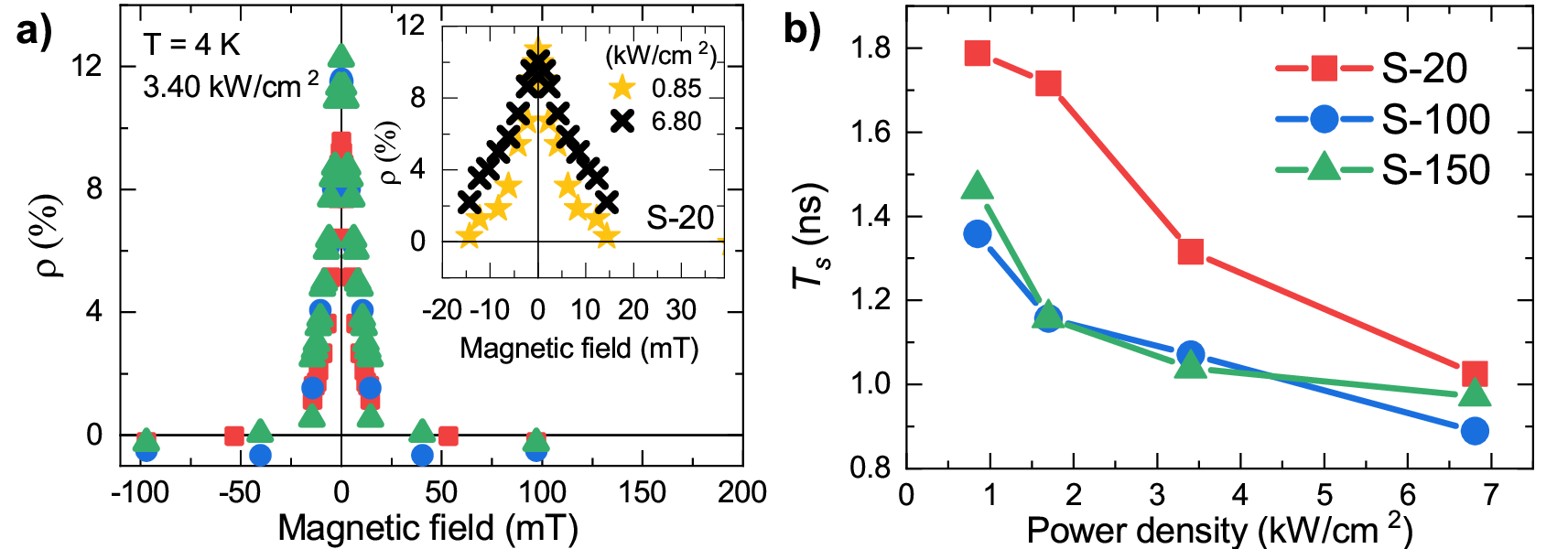}
  \caption{(a) Hanle curves of the $\mathrm{Ge_{0.91}Sn_{0.09}}$/Ge QW samples under an illumination of 3.40 $\mathrm{kW/cm^2}$. Data pertaining to sample S-20, S-100 and S-150 are shown as red squares, blue dots and green triangles, respectively. The inset shows Hanle curves of sample S-20 in a restricted region of the magnetic field at a power density of 0.85 (yellow stars) or 6.80 $\mathrm{kW/cm^2}$ (black crosses). All the data are mirrored to negative magnetic fields to show the characteristic Lorentzian shape pertaining to the Hanle effect. (b) Spin lifetime ($T_S$) derived from the Hanle curves measured at various illumination conditions. All the measurements were performed at a temperature of 4K using a right-handed circularly polarized laser with photon energy of 1.16 eV.}
  \label{fig2}
\end{figure*}

Given the intriguing possibility of addressing the emission of a single $\mathrm{Ge_{0.91}Sn_{0.09}}$ QW, we can hereafter utilize optical spin orientation \cite{Lampel68} and gather optical access to the fundamental spin-dependent properties of the 2DHG. 

To this purpose, we will evaluate the spin lifetime ($T_s$) at cryogenic temperatures, namely 4 K, by leveraging the recent extension of the Hanle effect to group IV materials achieved through magneto-PL.\cite{vitiello20} We therefore conducted experiments in a Voigt geometry, that is, the magnetic field (B) being perpendicular to the spin quantization axis. The latter is defined by the light propagation path being parallel to the growth direction, hence normal, in our configuration, to the QW plane.

Figure \ref{fig2}a shows that at a specific excitation power density of 3.4 kW/cm$^2$, all the samples demonstrate a sizable circular polarization degree ($\rho$). This is due to radiative recombination events involving spin-polarized carriers generated through the process of optical spin injection.\cite{Parsons69, Pezzoli12, decesari19}. In particular, at zero field, sample S-20 demonstrates a $\rho_0\equiv\rho(B=0)$ of about 9\%, whereas the polarization degree is higher for the other two samples and equal to $\sim 12\%$. This modification of $\rho_0$ can be likely ascribed to the inherent \textit{p}-MOD structure. The structural asymmetry that originates from the one-side modulation doping leads to the emergence of the so-called Rashba effect.\cite{Bychkov84, Rashba59} This acts as a perturbation to the spin ensemble, which causes a loss of the polarization degree, thereby shortening the spin relaxation time ($T_1$) and, in turns, decreases the observable $\rho_0$.\cite{Rossi22} This mechanisms becomes more pronounced when the strength of the effective electric field is increased by moving the doping layer closer to the QW as in S-20. 
%Supporting information additionally demonstrates that the temperature dependence of $\rho_0$ follows the known behavior for strained $\mathrm{Ge_{1-x}Sn_x}$ epitaxial layers.\cite{decesari19}

Notably, $\rho(B)$ of all the samples decays by increasing the strength of the field and gets completely washed out above $\sim 20$ mT. Such a well-defined behavior is fully consistent with the Lorentzian lineshape expected for the Hanle effect, i.e., $ \rho(B)=\rho_0 / \left[1+\left(\Omega T_S\right)^2\right]$,\cite{Dyakonov84} where the Larmor precession frequency $\Omega = (g \mu_B B)/\hbar$ depends on the strength of the magnetic field and the $g$ factor through the Bohr magneton ($\mu_B$) and the reduced Planck constant ($\hbar$). 

It is worth noting that a marked increase in the Hanle linewidth can be effectively induced by changing the power density of the optical pump from 0.85 to 6.80 kW/cm$^2$ (see the inset of Figure \ref{fig2}a). Using the predicted lineshape for the Hanle curves \cite{vitiello20} and the experimental g-factor of 1.4828 from Ref. \cite{decesari19}, we derived the change of $T_s$ upon the excitation power. These results are summarized in Figure \ref{fig2}b and readily manifests that a pump-induced shortening of the spin lifetime can occur in all the $\mathrm{Ge_{0.91}Sn_{0.09}}$/Ge QWs. We emphasize that the spin lifetime extracted from the Hanle curve arises from the intertwined contributions of the spin-relaxation time and the carrier lifetime ($\tau$), being $1/T_s = 1/T_1 + 1/\tau$. In particular, the observation of circularly polarized emission allows us to conclude that $T_1$ is practically longer than $\tau$. Consequently, any change in $T_s$ is mainly governed by modifications of the carrier lifetime.\cite{Parsons69,vitiello20} 

The type-II band alignment inhibits the coexistence of opposite charges within the same spatial layer. As a result, a reduced recombination probability and, in turn, a lengthening of the carrier lifetime can be expected in $\mathrm{Ge_{0.91}Sn_{0.09}}$/Ge QWs as compared to the bulk. Consistently, the values of the spin lifetime demonstrated in Figure \ref{fig2}b are in the ns range, which is markedly longer than the sub-ns regime reported by previous experimental works on epitaxial bulk-like films, provided that a Sn molar fraction similar to the one studied here is considered.\cite{vitiello20, Rogowicz21} The carrier lifetime obained via magneto-PL compares favorably also with recent results on SiGeSn/GeSn/SiGeSn QWS.\cite{Grant24} Moreover, it can be noticed that the longest lifetime shown in Figure \ref{fig2}b pertains to sample S-20. This QW owns the thinnest Ge spacer, i.e., the deepest confining potential, which ensures better bound states. More importantly, the large carrier lifetime of sample S-20 can also concur with the Rashba field to lower the value of $\rho_0$ as the polarization degree is ultimately dictated, under steady state conditions, by the ratio $T_1/\tau$.\cite{Parsons69, Pezzoli12, zutic04}   

The monotonic power-dependent decrease of $T_S$ and its marked effect observed in Figure \ref{fig2}b also stems from the type II nature of the band lineup in $\mathrm{Ge_{0.91}Sn_{0.09}}$/Ge QWs. The growth of the dense non-equilibrium distribution of carriers induced by enlarging the fluence of the optical pump progressively fills higher energy states of the QW, which are characterized by a larger penetration of the associated wave functions into the adjacent barrier layers. This process is of greater importance for QWs having a larger confinement energy and improves the otherwise small spatial overlap between the electrons and holes that are confined at the opposite sides of the hetero-interface. Eventually, this mechanism increments the oscillator strength governing the optical transitions and shortens the radiative recombination time as observed in Figure \ref{fig2}b.\cite{Zaitsev07,Baranowski12} In S-20 the increase of the power density causes indeed the most pronounced shortening of the spin lifetime as $T_S$ is almost halved being reduced from 1.8 ns until it equals the data of the other two QW samples at about 1 ns (see Figure \ref{fig2}b). 

\subsection{Optical control of the inverse Spin-Hall effect}

\begin{figure*}
  \includegraphics[width=\linewidth]{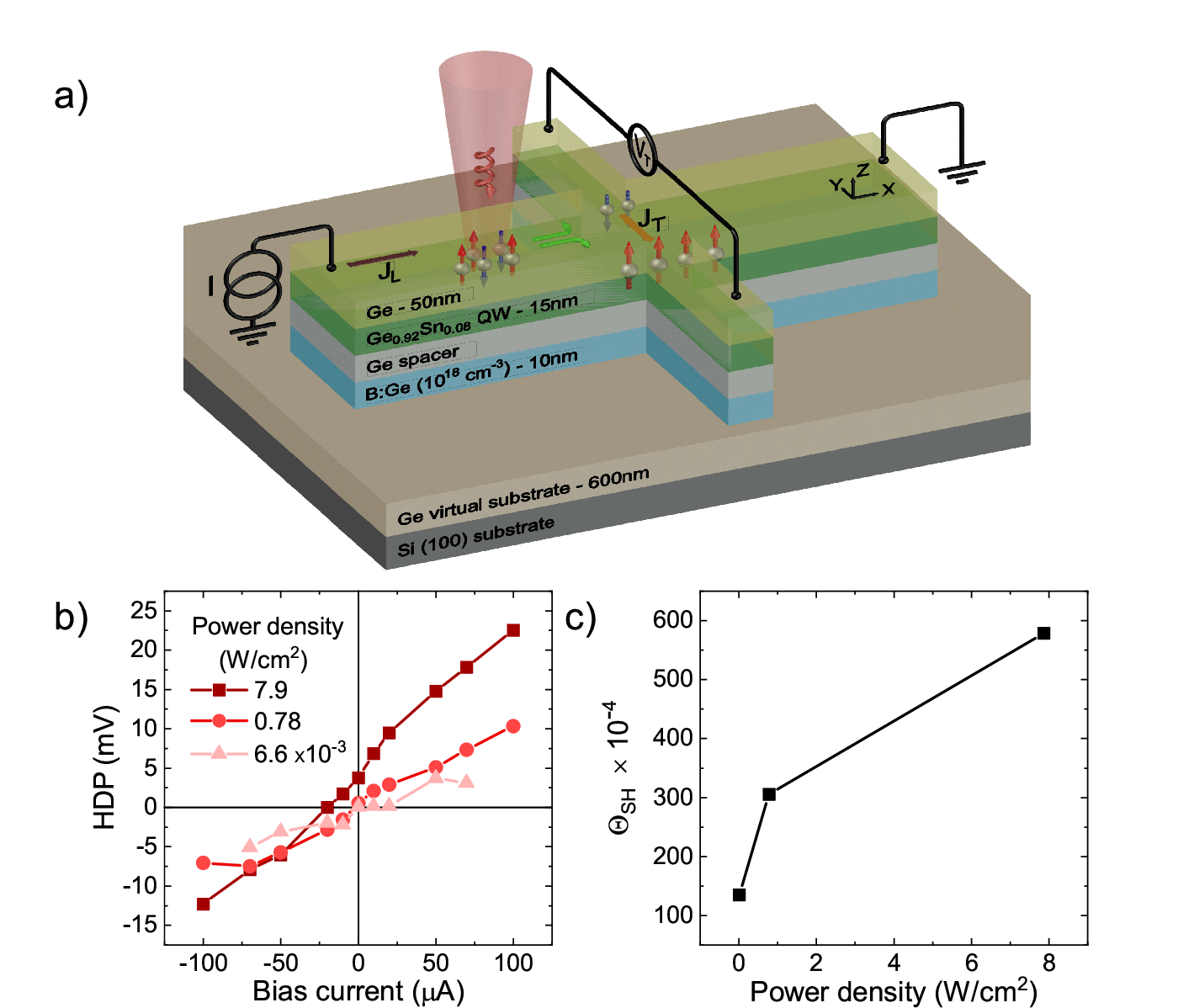}
  \caption{: (a) Sketch of the set-up used for the measurement of the inverse spin-Hall effect (ISHE). The transverse photovoltage ($V_T$) is measured across the side channel by a lock-in amplifier connected to a load resistance. (b) The difference between the ISHE voltage generated in the transverse channel of the sample S-150 under right- and left-handed circularly polarized excitation provides the helicity-dependent photovoltage, HDP, reported here as a function of the bias current for three different illumination power densities. (c) Spin-Hall angle derived from the HDP data as a function of illumination.}
  \label{fig3}
\end{figure*}

The presence of a structural inversion asymmetry, due to the one-sided MOD, offers the possibility to study the emergence of advanced phenomena in a 2DHG framework. Specifically, the Rasha-induced SOC, which is inherently associated with a spin texture in momentum space, can be leveraged to optically induce spin-to-charge interconversion via ISHE. 

Figure \ref{fig3}a shows the experimental geometry utilized to demonstrate this process in the type II $\mathrm{Ge_{0.91}Sn_{0.09}}$/Ge heterostructure. A circularly polarized laser beam is steered along the normal (hereafter versor $\hat{z}$) to the  Hall-bar surface, i.e., the QW plane. By labeling the Bloch states according to the total angular momentum ($J$) and its projection onto the $z$ axis ($J_z$), it can be easily shown that, by virtue of the selection rules, interband transitions generate electron and heavy hole ensembles holding the same density and the same out-of-plane spin orientation.\cite{Dyakonov84, zutic04} The latter can be either parallel or antiparallel to the direction of light propagation and is defined by the helicity of the laser. For example, a right-handed $\sigma^+$ pump promotes electrons to $\left|J; J_z\right\rangle = \left|1/2; -1/2\right\rangle$ conduction states, while injecting heavy holes into $\left|3/2; -3/2\right\rangle$ states.  

The application of an electric field along the longitudinal (main) channel of the device will then cause the drift within the QW plane of the photogenerated spin-polarized carriers. The resulting electron and heavy hole spin currents stem from the anticommutator of velocity and spin \cite{Bernevig05} and flow in opposite directions, following the relative motion of the charged species. It should be noted, however, that these two contributions to the overall spin current will not cancel out because of the different magnitude of $J_z$. Morevover, photoexcitation from light hole states, which is accessible in our experimental conditions, results in the same spin orientation of the HHs, while reducing the overall electron spin polarization in the conduction band. This ultimately ensures that holes drag longitudinally a net spin angular momentum. Such mechanisms, in turns, generates through spontaneous ISHE a genuine flow of charges and a voltage drop at the transverse (side) channel, $\mathrm{V_T}$, that can be measured across a load resistor via standard lock-in techniques (see Methods for details).\cite{okamoto14, liu18} The sign of $\mathrm{V_T}$ is indeed consistent with positive charge carriers involved in the transport process.

To correct for spurious effects and to determine the unique spin-related components of the electro-optical signal, we single out the helicity-dependent photovoltage (HDP), that is, the difference between the ISHE voltage, $\mathrm{V_T^+}$ and $\mathrm{V_T^-}$, generated in the resistive side arm under right- and left-handed circularly polarized excitation, respectively. Fig. \ref{fig3}b summarizes the HDP obtained for three different laser fluences by sweeping the current in the $\pm 100$ $\mu$A interval. As expected for the ISHE (see Methods), HDP demonstrates a sign inversion when the current direction is reversed. Moreover, a linear dependence of the ISHE photovoltage on the longitudinal current can be observed over a wide range of the excitation conditions. All these findings provide a compelling demonstration of the spin-to-charge interconversion occurring in the $p$-MOD structure. 

It should be noted that when the power density is increased to 8 $\mathrm{W/cm^2}$ (dark-red squares in Fig. \ref{fig3}b), there still exists a residual HDP of approximately 5 mV even in complete absence of the longitudinal bias. This negligible offset can be regarded as the onset of laser-induced  heating effects. The HDP measured as a function of the longitudinal current, in combination with the value of the spin polarization obtained by the optical spin orientation experiments, enabled us to finally extract the spin-Hall angle $\theta_{SH}$, i.e., the figure of merit for the efficiency of the spin-to-charge interconversion process (see Methods). As shown in Figure \ref{fig3}c, the strength of the SHE increases monotonically with the power density going from $130 \times 10^{-4}$ to almost $600\times 10^{-4}$, when the excitation is increased from $7\times 10^{-3}$ to 8 $\mathrm{W/cm^2}$. 

\begin{table*}
 \caption{Spin-Hall angle $\theta_{SH}$ for different materials. The temperature at which the spin-Hall angle is extracted is reported along with the bibliographic reference and the relevant charge species (electron: e, or hole: h) involved in the process. \label{tab:1}}
%  \begin{tabular}[]{@{}lllll@{}}
\begin{tabular}{lllll}
    \hline
    Material & $\theta_{SH}$ & Temperature (K) & charge & Reference\\
    \hline
%       Al  & 0.4-1  & 4.2  & \cite{valenzuela06} \\
%		Pd  & 64  & - & \cite{Mosendz10} \\
%		Ta  & 2000 & 300 & \cite{Liu12} \\
		GaAs  & 5-200 $\times 10^{-4}$ & 300 & e & \cite{okamoto14} \\
		Si  & 1 $\times 10^{-4}$ & 300 & e & \cite{ando12} \\
        Ge  & 2-11 $\times 10^{-4}$ & 20-300   & e & \cite{Rojas13,Bottegoni17}  \\
		$\mathrm{Ge_{0.95}Sn_{0.05}}$  & $4\times 10^{-5}$ - $3\times 10^{-1}$  & 300  & e & \cite{marchionni21}  \\
		$\mathrm{Ge_{0.91}Sn_{0.09}}$/Ge QW  & 130 - 600 $\times 10^{-4}$  & 4  & h & This work  \\
    \hline
  \end{tabular}
\end{table*}

A quick glance at table \ref{tab:1} enables us to conclude that 2DHG in staggered $\mathrm{Ge_{0.91}Sn_{0.09}}$/Ge stands out as a spintronic candidate, outperforming other heterostructures, particularly those consisting of group IV semiconductors. The $\mathrm{Ge_{0.91}Sn_{0.09}}$/Ge QW demonstrates indeed a 10$x$-60$x$ efficiency increase compared to bulk Ge, and, under comparable pump conditions, a two-fold improvement with respect to $\mathrm{Ge_{0.95}Sn_{0.05}}$ epilayers.\cite{marchionni21} Such distinct advantage is in line with (i) the enlarged SOC caused by alloying Ge with the heavier element Sn, and (ii) the Rashba-induced spin-splitting stemming from the structural inversion asymmetry.

Besides an efficient spin-to-charge conversion, Figure \ref{fig3}c demonstrates the unrivaled benefit offered by the type-II $\mathrm{Ge_{0.91}Sn_{0.09}}$/Ge QW, that is, the practical control of $\theta_{SH}$ through optical excitation. The magneto-optical investigations, discussed previously, have shown that the laser fluency systematically modifies the band-bending and state filling because of the pump-induced reinforcement of the potential drop across the hetero-interface. According to $k\cdot p$ theory, the field gradient enters into the Rashba coefficient $\alpha_R$, which couples in the Rashba Hamiltonian the spin Pauli matrices to the momentum and determines the strength of the SOC exerted on the 2DHG confined in the valence band of the QW. \cite{Winkler03} It is thus reasonable to conclude that optical pumping in staggered heterostructures can modify, by means of carrier accumulation and state filling effects, the spin-orbit Hamiltonian that governs spin splitting and spin transport, ultimately introducing a viable and unique approach to engineer the Rashba physics occurring in a 2DHG, as shown by Figure \ref{fig3}c.

Finally, it is instructive to connect these ISHE results with the theory of the injection of spin currents at Rashba interfaces, as  established within the framework of the inverse Edelstein effect (IEE).\cite{edelstein90, Ka14} By neglecting, to a first approximation, the asymmetric confining potential of the MOD-QW, we can assume that the photogenerated spins are homogeneously distributed within the illuminated volume, defined by the circular spot of the laser (diameter $d \simeq 50 \mathrm{\mu m}$) and the thickness of the $\mathrm{Ge_{1-x}Sn_x}$ QW. The latter being smaller than the light penetration depth. Under such constraints and through the action of the longitudinal electric field, we can assume the establishment of a 3D flux density, $j_s$, given by the flow of the spin-polarized holes along the path length $d$. A transverse charge current, $j_c$, eventually manifests itself within the 2D Rashba-split hole gas. Such a conversion process is conceptually equivalent to the well-known IEE occurring in spin pumping experiments, where electrons are injected from a ferromagnetic contact into a nonmagnetic layer, whose thickness is smaller than the spin diffusion length, hence:\cite{sanchez13,Rojas-Sanchez16}

\begin{equation}
	j_c = \frac{d}{2} \theta_{SH} j_s
	\label{eq1}
\end{equation}

Similarly, this relationship can be recast in terms of the Rashba coefficient as:\cite{sanchez13, Rojas-Sanchez16}

\begin{equation}
	j_c = \alpha_R \frac{\tau^h_S}{\hbar} j_s
	\label{eq2}
\end{equation}

By assuming a spin-relaxation time for holes ($\tau^h_S$) between 2 and 20 ps \cite{tai21, Lange12} and by combining Eqs \ref{eq1} and \ref{eq2}, we can estimate for p-MOD $\mathrm{Ge_{0.91}Sn_{0.09}}$ QWs a linear coefficient $\alpha_R$ in the 0.1-5 meV ${\mathrm{\AA}}$ range. This appears to be compatible with the value of the Rashba coefficient derived within the Kane model, e.g., for electrons\cite{Winkler03}. Indeed, $\alpha_R = 0.5$ meV ${\mathrm{\AA}}$ is obtained using the experimental energy gap ($\sim 0.6$ eV), the split-off (0.44 eV) and the electric field (25 kV/cm) taken from band profile simulations \cite{Birner07} and evaluating the $k\cdot p$ matrix element ($\sim$-1.84 eV ${\mathrm{\AA}}$) at the Sn molar fraction of the QW according to Ref. \cite{song19}.

It should be noted that, within the Luttinger formalism, the SOC Hamiltonian of HHs should solely retain the Rashba coefficient of the $k$-cubic component.\cite{Murakami04, Bernevig05} Yet, refined theoretical investigations have established the existence of a Rashba term that is linear in $k$ and originates from bulk inversion asymmetry.\cite{Winkler03} Although this Hamiltonian contribution is small and usually negligible, it can be expected to become important in the low-density regime, where the relative weight of the $k$-cubic Rashba is strongly reduced.\cite{Winkler03} While the linear Rashba coefficient for HH has not been experimentally and theoretically addressed in $\mathrm{Ge_{1-x}Sn_{x}}$ alloys thus far, we can nevertheless find reassurance of the accuracy of our derivation by the satisfactory comparison with recent theoretical predictions set out for HHs in QWs consisting of elemental Ge.\cite{Xiong22, Rodriguez-Mena23} Even for Ge QWs, indeed, the experimental determination of the linear $\alpha_R$ remains beyond the grasp of transport experiments, whose resolution is practically limited to higher-order cubic terms.\cite{Moriya14, Morrison14, tai21} 

The informative role of optical investigations on the k-linear Rashba coefficient therefore unlocks intriguing properties of 2DHGs pertaining to group IV heterostructures and holds the potential to stimulate future studies on the unconventional SOC at type-II interfaces.

\section{Conclusion}\label{sec11}

We have introduced an all-optical investigation of \emph{p}-type MOD $\mathrm{Ge_{0.91}Sn_{0.09}}$/Ge QWs. These heterostructures are found to remarkably host a staggered band-edge alignment, which offers an original platform for the investigation of the rich spin physics emerging in a Rashba-split hole gas. We applied, within the context of the well-consolidated continuous-wave PL, the Hanle effect to gather information on the carrier dynamics. Specifically, the magneto-optical investigation demonstrated that quantum confinement in a type-II heterostructure results in a spatial indirect nature of the excitonic recombination, which manifests itself as a few-ns-long carrier lifetime.
Moreover, the structural inversion asymmetry encoded in the MOD epitaxial architecture lifts the spin degeneracy of the band structure without requiring external magnetic fields. The resulting spin texture is enhanced by the tunable SOC introduced by alloying Ge with Sn and by the additional degree of freedom offered by the low dimensionality. Such properties constitute an interesting playground to address intriguing phenomena such as spin-to-charge interconversion via ISHE. Our experiments demonstrate that a 2DHG forming at a staggered $\mathrm{Ge_{1-x}Sn_x}$/Ge heterojunction permits the optical reconfiguration of the Rashba Hamiltonian and ensures full control over the the spin-Hall angle. The ability to create a robust spin current and spin-to-charge conversion through a contact-less optical approach in a 2DHG is indeed intriguing and highly desirable because it offer the unprecedented access to the Rashba term in the HH SOC Hamiltonian that is linear in the in-plane wave vector. Specifically, the application of $\mathrm{Ge_{0.91}Sn_{0.09}}$/Ge QWs can pave the way to future explorations of electro-optical manipulation of spins in quantum technologies requiring SOC control with the notable addition of a sizable spin-photon interaction. 

\section{Methods}\label{sec12}

\textbf{Sample growth and device processing} Three samples are grown in a ASM Epsilon 2000 industrial-type reduced-pressure chemical vapor deposition system. The precursor materials are $\mathrm{SnCl_4}$ and $\mathrm{Ge_2H_6}$ with an overall $\mathrm{H_2}$ atmosphere in the system. To avoid segregation, a common problem in systems with Sn $> 7$\%, a low growth temperature of approximately 270 $\mathrm{^o}$C is used during the process. The Ge spacer layer thickness (h) was changed between 20, 100 and 150 nm. On each sample, we patterned Hall-bars with a Heidelberg Instrument $\mu$PG 101 laser writer. Contact metallization was performed via sputtering deposition of 100-nm-thick Pt layer, while a wet etch in $\mathrm{H_2O_2:HCl:H_2O}$ $(1:1:20)$ was used to finalize the mesa. For the ISHE experiments, a dedicated set of Hall-bars with a width $w$ of 200 $\mu$m and a length $L$ of 5 mm was fabricated, this requirement is necessary due to the laser spot dimension.

\textbf{Optical measurements} PL spectra were measured at 4K under the excitation of a right-handed circularly-polarized $\mathrm{Nd:YVO_4}$ laser at 1064 nm (1.165 eV). The laser spot diameter on the sample surface was approximately 50 $\mu$m.\cite{pezzoli16} The polarization of the PL was analyzed using a linear polarizer and a quarter-waveplate. The degree of circular polarization, which is a measure of the different intensity between the right- $(I^+)$ and left- $(I^-)$ handed circularly polarized components of the PL was determined by performing a full Stokes analysis (see Refs. \cite{Pezzoli12, Pezzoli13} and therein for details). The measurement of the PL intensity was conducted by coupling a monochromator to a long-wavelength single channel (In,Ga)As photodiode using a standard lock-in technique. Electrical bias was applied to the device through a Keithley source measure unit.

\textbf{Inverse Spin Hall measurements} We performed photovoltage measurements by connecting the side channel of the Hall bar in series to a load resistance, $R_L$, of 10 k$\Omega$, while measuring the voltage drop across the latter via a lock-in amplifier. For the experiment, we then connected the main channel to a current generator. The parasitic transverse electromotive force, potentially caused by illumination-induced thermal gradients, has been minimized in the experiment by locating the optical beam in between the two side electrodes. The polarization-induced changes of the transverse voltage is thus:\cite{okamoto14, liu18}

\begin{equation}
	\Delta V_T = V_T^+ - V_T^- = \theta_{SH} r w J_L (P^+ - P^-)
	\label{eqn}
\end{equation}

where $\theta_{SH}$, $r$ and $J_L$ are the spin Hall angle, the resistivity of the sample, the longitudinal current density (the longitudinal current over the cross-sectional area), while $P^+$ and $P^-$ are the carrier spin polarizations under $\sigma^+$ and $\sigma^-$ excitation, respectively. Knowing the circular polarization degree through optical spin orientation measurements, $\rho$, we can retrieve the carrier spin polarization $P^+ = 2 \rho$. From the circuit, we extract the helicity-dependent photovoltage, HDP, generated across the sample as:

\begin{equation}
	HDP=\frac{R_S}{R_L}\Delta V_T
	\label{eqn1}
\end{equation}

where $R_L$ is the load resistance and $R_S$ is the resistance of the sample. Due to the very low illumination power density (at least three orders of magnitude with respect to the PL measurements), $R_S$ results from the fit of I-V curve under dark condition and accounts to 210 $k\Omega$ at 4 K. 

\section{Acknowledgments}\label{sec14}
The authors would like to acknowledge L. Lonobile and B.M. Ferrari for technical assistance with the measurements. F.P. acknowledges support by the Air Force Office of Scientific Research under the award number FA8655-22-1-7050.

\end{document}